\documentclass[fleqn,usenatbib]{mnras}

\usepackage{newtxtext,newtxmath}

\usepackage[T1]{fontenc}
\usepackage{ae,aecompl}
\usepackage{multicol}

\usepackage{enumerate}
\usepackage[normalem]{ulem}
\usepackage[usenames]{color}

\usepackage{graphicx}	
\usepackage{amsmath}	
\usepackage{amssymb}	

\newcommand{\voro}{{\sevensize VORO++}}
\newcommand{\astrobear}{{\sevensize ASTROBEAR}}

\newcommand{\gasoline}{{\sevensize GASOLINE}}
\newcommand{\arepo}{{\sevensize AREPO}}
\newcommand{\changaMM}{{\sevensize MANGA}}
\newcommand{\changa}{C{\sevensize HA}NG{\sevensize A}}

\newcommand{\mesa}{{\sevensize MESA}}
\newcommand{\be}{\begin{eqnarray}}
\newcommand{\ee}{\end{eqnarray}}
\newcommand{\rsun}{\ensuremath{\textrm{R}_{\odot}}}
\newcommand{\msun}{\ensuremath{\textrm{M}_{\odot}}}

\newcommand{\grad}{\ensuremath{\boldsymbol{\nabla}}}
\newcommand{\vel}{\ensuremath{\boldsymbol{v}}}
\newcommand{\velCM}{\ensuremath{\boldsymbol{v}_{\rm CM}}}

\newcommand{\ddt}[1]{\ensuremath{\frac{\partial #1}{\partial t}}}
\newcommand{\ddx}[1]{\ensuremath{\frac{\partial #1}{\partial x}}}
\newcommand{\state}{\ensuremath{\boldsymbol{\mathcal{U}}}}
\newcommand{\charge}{\ensuremath{\boldsymbol{U}}}

\newcommand{\flux}{\ensuremath{\boldsymbol{\mathcal{F}}}}
\newcommand{\fluxV}{\ensuremath{\boldsymbol{F}}}
\newcommand{\source}{\ensuremath{\boldsymbol{\mathcal{S}}}}
\newcommand{\sourceV}{\ensuremath{\boldsymbol{S}}}

\newcommand{\pdiff}{\ensuremath{\epsilon_{p}}}
\newcommand{\pcont}{\ensuremath{P_{\rm{con}}}}
\newcommand{\facecentroid}{\ensuremath{\tilde{\boldsymbol{r}}}}
\newcommand{\cellcentroid}{\ensuremath{\boldsymbol{c}}}
\newcommand{\force}{\ensuremath{\boldsymbol{F}}}
\newcommand{\accel}{\ensuremath{\boldsymbol{a}}}

\newcommand{\normal}{\ensuremath{\hat{\boldsymbol{n}}}}

\newcommand{\facev}{\ensuremath{\boldsymbol{\tilde{w}}_{ij}}}
\newcommand{\facer}{\ensuremath{\boldsymbol{\tilde{r}}_{ij}}}

\newcommand{\dt}{\ensuremath{\delta t}}

\newcommand{\Etoti}{\ensuremath{E_{\rm tot,i}}}

\newcommand{\munbound}{\ensuremath{m_{\rm g,u}}}
\newcommand{\funbound}{\ensuremath{f_{\rm unb}}}

\newcommand{\lp}[1]{\textrm{\color{red} #1}}

\renewcommand{\sout}[1]{}
\renewcommand{\lp}[1]{#1}

\title[Moving Boundary Conditions in Hydrodynamics]{Moving and Reactive Boundary Conditions in Moving-Mesh Hydrodynamics}

\author[]{ Logan J. Prust$^1$\thanks{LJP: ljprust@uwm.edu}
\\
$^1$ Department of Physics, University of Wisconsin-Milwaukee, 3135 North Maryland Avenue, Milwaukee, WI 53211, USA\\
}

\date{Accepted XXX. Received YYY; in original form ZZZ}

\pubyear{2019}

\begin{document}
\label{firstpage}
\pagerange{\pageref{firstpage}--\pageref{lastpage}}
\maketitle

\begin{abstract}
We outline the methodology of implementing moving boundary conditions into the moving-mesh code \changaMM. The motion of our boundaries is reactive to hydrodynamic and gravitational forces. We discuss the hydrodynamics of a moving boundary as well as the modifications to our hydrodynamic and gravity solvers. Appropriate initial conditions to accurately produce a boundary of arbitrary shape are also discussed. Our code is applied to several test cases, including a Sod shock tube, a Sedov-Taylor blast wave and a supersonic wind on a sphere. We show the convergence of conserved quantities in our simulations. We demonstrate the use of moving boundaries in astrophysical settings by simulating a common envelope phase in a binary system, in which the companion object is modeled by a spherical boundary. We conclude that our methodology is suitable to simulate astrophysical systems using moving and reactive boundary conditions.
\end{abstract}

\begin{keywords}
methods: numerical --- hydrodynamics
\end{keywords}



\section{Introduction}\label{sec:intro}

Numerical simulations have played a crucial role in understanding the hydrodynamics of many astrophysical phenomena, including gas, stars, discs, galaxies and large-scale structure. Two methodologies are commonly employed to solve the hydrodynamic equations: smooth particle hydrodynamics (SPH) and grid-based solvers. SPH codes \lp{\sout{conserve momentum}have excellent conservation properties} but resolve shocks poorly due to their smoothing nature. On the other hand, grid-based solvers are superior at modeling shocks due to the use of Godunov schemes. \lp{\sout{However, grid effects can lead to numerical diffusion and poor conservation properties.}However, they suffer from a lack of Galilean invariance and errors in angular momentum conservation. Furthermore, even with adaptive mesh refinement, grid-based solvers encounter difficulty with highly-refined regions that move with large velocities relative to the mesh. This is due to the challenges associated with adapting the mesh to the anticipated motion of the system \citep{2010MNRAS.401..791S}.}

In recent years, a hybrid of these methods has been developed in an attempt to capture the best characteristics of both. This is the arbitrary Lagrangian-Eulerian (ALE) scheme, and software that use ALE schemes are known as moving-mesh codes \citep[for a review see][]{alebook}. In an ALE scheme, the mesh moves along with the fluid, combining the superior shock-capturing of grid-based solvers with the conservation properties of SPH. \citet[][hereinafter S10]{2010MNRAS.401..791S} described one such scheme that has proven successful, which is implemented in the code \arepo. The scheme constructs an unstructured mesh from an arbitrary distribution of points using a Voronoi tessellation \citep{1992stca.book.....O}. This guarantees that the mesh will be well-defined, unique and continuously deformable; thus, finite volume methods can be applied in a manner similar to that of an Eulerian code. In addition, the lack of Galilean invariance in Eulerian codes is rectified if the mesh cells move along with the local flow. It has also been argued that ALE schemes are superior at capturing boundary layer instabilities such as Kelvin-Helmholtz instabilities (S10), though \citet{2016mnras.455.4274l} cautioned that numerical noise can masquerade as solutions of such instabilities. Although \arepo\ was originally developed with cosmological simulations in mind \citep[see for instance][]{2014MNRAS.444.1518V}, it has also been used in a number of problems including stellar mergers \citep{2015ApJ...806L...1Z}, tidal disruption events \citep{2019MNRAS.487..981G} and common envelope evolution \citep{2016ApJ...816L...9O}.

A moving-mesh hydrodynamic solver has also been developed for the N-body simulation code \changa\ (Charm N-body GrAvity solver) \citep{jetley2008,jetley2010,2015comac...2....1m}. This moving-mesh solver is known as \changaMM, which is described in detail by \citet[][hereinafter CWQ17]{Chang+17}. In \citet[][hereinafter PC19]{2019MNRAS.486.5809P}, we incorporated individual time-steps into \changaMM\ as well as the equations of state implemented in the open source stellar evolution code \mesa\ \citep{2011ApJS..192....3P,2013ApJS..208....4P,2015ApJS..220...15P,2018ApJS..234...34P,2019ApJS..243...10P}. We used these tools to simulate common envelope evolution (CEE) in binary star systems.

In fluid dynamics it is often desirable to set boundary conditions on a surface. For example, in modeling the flow around an airfoil the edges of the simulation box can be treated as inflow or outflow boundaries and the airfoil itself as a reflective boundary, which allows no matter to pass through it. Indeed, \citet{2019arXiv190601813W} use a 2-D ALE scheme with moving boundaries to model a pitching NACA 0012 airfoil. Immersed boundary methods can also be useful, especially when the relative displacement of the boundaries is large \citep{2019arXiv191009315H}.

Such schemes are common in fields such as aerodynamics, but also have interesting and useful applications in astrophysics. For example, \citet{2019arXiv191008027C} use two separate boundary conditions in their simulations of asymptotic-giant-branch (AGB) binaries using the adaptive mesh refinement code \astrobear. On a cylindrical surface surrounding the binary, they set the fluid velocity to be supersonic in the radially outward direction, ensuring that no information propagates into the cylinder from the outside. They also set an inner boundary condition beneath the photosphere of the AGB star requiring that the radial fluid velocity oscillate in time, which approximates stellar pulsations \citep{1988ApJ...329..299B}. Additionally, inflow conditions can be useful in astrophysical problems where accretion is present, such as in the growth of intermediate mass black holes in the early universe \citep[for example,][]{2019MNRAS.483.2031T}. However, due to the difficulty of implementing moving boundaries, such phenomena are typically modeled using static boundary conditions even in cases where moving boundaries are a more natural fit to the system in question.

The implementation of boundary conditions into Eulerian grid-based solvers, which we discuss in section \ref{sec:hydro}, is straightforward. However, because Eulerian grid codes have a static mesh, it is at best very difficult to have a boundary which moves with respect to the mesh. On the other hand, moving-mesh codes are ideal for moving boundaries since the mesh-generating points can move with respect to one another. In ALE schemes, the velocities of the mesh-generating points are typically taken to be equal or similar to the fluid velocity to gain the advantages offered by a Lagrangian scheme. However, the mesh-generating points can in principle move at any velocity relative to the fluid.

S10 carried out a simple 2-D simulation using moving boundaries in \arepo\ in which a curved boundary (thought of as a ``spoon'') stirs a fluid. They generate their boundary using two sets of points on either side of and equidistant from the desired boundary surface (see their Fig. 38), on which they set reflective boundary conditions. They also require that both sets of points move together as a rigid body, and move their spoon at a constant speed along a pre-determined circular path.

We adopt a similar strategy for our implementation of moving boundaries into \changaMM, requiring the motion of the cells adjacent to the boundary to be identical to the motion of the boundary itself. However, we also set the velocity of the boundary such that its motion is reactive to external forces. The aerodynamic drag force can be computed by integrating the gas pressure over the surface of the boundary, and the gravitational force by integrating over the volume of the boundary.

We have organized this paper as follows. In section \ref{sec:previous}, we summarize the algorithm of \changaMM\ prior to the implementation of moving boundary conditions. We discuss the changes to the hydrodynamic and gravity solvers to implement moving boundaries in section \ref{sec:hydro}. In section \ref{sec:test}, we run simulations of four test cases: a supersonic wind, a Sod shock tube, a Sedov-Taylor explosion and a common envelope phase in a binary star system. We discuss our results and possible future directions in section \ref{sec:discussion}.

\section{Previous Algorithm}\label{sec:previous}

The ALE algorithm that is implemented in \changaMM\ is briefly summarized as follows. We refer the reader to CWQ17 and PC19 for detailed discussions. \changaMM\ solves the Euler equations, which written in conservative form are
\be
\ddt{\rho} + \grad\cdot\rho\vel &=& 0 \label{eq:continuity},\\
\ddt{\rho\vel} + \grad\cdot\rho\vel\vel + \grad P &=& -\rho\grad\Phi\label{eq:momentum}
\ee
\begin{flushleft}
and
\end{flushleft}
\be
\ddt{\rho e} + \grad\cdot\left(\rho e + P\right)\vel &=& -\rho\vel\cdot\grad\Phi\label{eq:energy},
\ee
where $\rho$ is the density, $\vel$ is the fluid velocity, $\Phi$ is the gravitational potential, $e= \epsilon + v^2/2$ is the specific energy, $\epsilon$ is the internal energy and $P(\rho, \epsilon)$ is the pressure.  Equations (\ref{eq:continuity}) to (\ref{eq:energy}) can be written in a compact form by introducing a state vector $\state=(\rho, \rho\vel, \rho e)$:
\be
\ddt{\state} + \grad\cdot\flux = \source\label{eq:state},
\ee
where $\flux=(\rho\vel, \rho\vel\vel + P, (\rho e + P)\vel)$ is the flux function and $\source = (0, -\rho\grad\Phi, -\rho\vel\cdot\grad\Phi$) is the source function. To solve equation (\ref{eq:state}), we adopt the same finite volume strategy as S10. For each cell, the integral over the volume of the $i$th cell $V_i$ defines the charge of the $i$th cell to be
\be
\charge_i = \int_{V_{i}} \state dV = \state_i V_i.
\ee
As do S10, we then use Gauss' theorem to convert the volume integral over the divergence of the flux in equation (\ref{eq:state}) to a surface integral
\be
\int_{V_{i}} \grad\cdot\flux dV = \int_{\partial V_{i}} \flux\cdot\normal dA,
\ee
where $\partial V_{i}$ is the boundary of the cell. We now take advantage of the fact that the volumes are Voronoi cells with a finite number of neighbours to define an integrated flux
\be
\sum_{j \in {\rm neighbors}} \fluxV_{ij} A_{ij} = \int_{\partial V_{i}} \flux\cdot\normal dA,
\ee
where $\fluxV_{ij}$ and $A_{ij}$ are the average flux and area of the common face between cells $i$ and $j$.
The discrete time evolution of the charges in the system is given by
\be
\charge_i^{n+1} = \charge_i^n + \delta t \sum_j \hat{\fluxV}_{ij} A_{ij} + \Delta t\sourceV_i, \label{eq:time evolution}
\ee
where $\hat{\fluxV}_{ij}$ is an estimate of the half time-step flux between the initial $\charge_i^n$ and final states $\charge^{n+1}_i$ and $\sourceV_i = \int_{V_{i}} \source dV$ is the time-averaged integrated source function.

We estimate the flux $\hat{\fluxV}_{ij}$ across each face as follows.
\begin{enumerate}[(i)]
 \item Use the gradient estimates at the initial time-step to predict the half time-step cell-centred values.
 \item Drift the cells a half time-step and rebuild the Voronoi tessellation at the half time-step.
\item Estimate the half time-step state vector (in the rest frame of the moving face) at the face centre (\facer) between the neighboring $i$ and $j$ cells by linear reconstruction. \label{item:half step}
 \item Estimate the (half time-step) velocity $\facev$ of the face, following S10, and boost the state vector from the ``lab'' frame (the rest frame of the simulation box) to the rest frame of the face to find the flux along the normal of the face. I.e., in the direction from $i$ to $j$.
 \item Estimate the flux $\hat{\fluxV}_{ij}$ across the face using an HLLC or HLL approximate Riemann solver implemented following \citet{toro2009riemann}.
 \item Boost the solved flux back into the ``lab'' frame.
\end{enumerate}
We can then use the estimated fluxes to time-evolve the charges $\charge_i$ following equation (\ref{eq:time evolution}) using the full time-step $\dt$ and apply changes owing to the source terms.

To conclude our summary of \changaMM, we outline the main computation loop as a whole.

\begin{enumerate}[(i)]

    \item We determine a valid Voronoi tessellation of the mesh-generating points using the \voro\ library.
    
    \item Using the volume of the Voronoi cell and integral quantities, $\charge$, the conserved and primitive variables are determined. The local gradients and half time-step conserved variables are calculated.
    
    \item The flux across the faces is calculated as discussed above.
    
    \item The new state $\charge'$ is determined from the fluxes. New velocities for the mesh generating points are determined. The mesh generating points drift by a full time-step.
    
\end{enumerate}

\section{Hydrodynamics of a Moving Boundary}\label{sec:hydro}

We now discuss our changes to the \changaMM\ algorithm to implement moving boundary conditions. The treatment is restricted to reflective boundaries, though the extension to inflow or outflow conditions is straightforward.

\subsection{Flux Across a Reflective Boundary}\label{sec:flux}

Let us consider a face on which we would like to set a boundary condition and boost our state vectors and flux functions into the rest frame of the face. If we rotate the face such that its normal lies along the $x$-direction, as is done in \changaMM, then we can write the flux normal to the face as
\be
\flux_{x} = (\rho v_{x},~\rho v_{x}^{2} + P,~\rho v_{x} v_{y},~\rho v_{x} v_{z},~(\rho e + P) v_{x} ). \label{eq:flux_x}
\ee
Here we have expanded the momentum flux tensor $\rho\vel\vel$ into its three components. The flux of mass, momentum and energy that pass through the face are determined by the values of the state vector $\state_{\rm{L}}$, $\state_{\rm{R}}$ and the flux function $\flux_{x,\rm{L}}$, $\flux_{x,\rm{R}}$ on either side of the face. We see that under the transformation $v_{x} \rightarrow -v_{x}$ all terms in $\flux_{x}$ change sign with the exception of the $x$-momentum flux $\rho v_{x}^{2} + P$. Conversely, when this transformation is applied to the state vector $\state=(\rho, \rho v_{x}, \rho v_{y}, \rho v_{z}, \rho e)$, the \textit{only} term that changes sign is the $x$-momentum $\rho v_{x}$. Consider now the Harten-Lax-van Leer (HLL) approximate Riemann solver \citep{1988SJNA...25..294E}, which computes the flux across the face as
\be
\hat{\fluxV}_{\rm{HLL}} = \frac{ \alpha_{+}\flux_{x,\rm{L}} + \alpha_{-}\flux_{x,\rm{R}} - \alpha_{+}\alpha_{-}(\state_{\rm{R}}-\state_{L}) }{ \alpha_{+}+\alpha_{-} },
\ee
where $\alpha_{\pm} = \rm{max}[0, \pm\lambda_{\pm}(\state_{\rm{L}}), \pm\lambda_{\pm}(\state_{\rm{R}})]$, the maximum and minimum eigenvalues of the Jacobians of each state are $\lambda_{\pm} = v_{x} \pm c_{\rm{s}}$ and $c_{\rm{s}}$ is the sound speed. We now impose the condition that the state vector and flux on either side are equal except for the fluid velocities normal to the face, which are opposite: $v_{x} = v_{x,\rm{L}} = -v_{x,\rm{R}}$. That is,
\be
\flux_{x,\rm{R}}(v_{x}) = \flux_{x,\rm{L}}(-v_{x}) \quad\text{and}\quad \state_{\rm{R}}(v_{x}) = \state_{\rm{L}}(-v_{x}). \label{eq:reflective}
\ee
This implies $\alpha_{+} = \alpha_{-} = c_{\rm{s}}+|v_{x}|$ and simplifies the HLL flux to the form
\be
\hat{\fluxV}_{\rm{HLL}} = \frac{ \flux_{x,\rm{L}}(v_{x}) + \flux_{x,\rm{L}}(-v_{x}) - \alpha_{+}[\state_{\rm{L}}(-v_{x})-\state_{\rm{L}}(v_{x})] } {2}.
\ee
If we now plug in $\flux$ and $\state$ in component form as in (\ref{eq:flux_x}), we obtain
\be
\hat{\fluxV}_{\rm{HLL}} = (0,~\rho v_{x}^{2}+P+\rho v_{x}(c_{\rm{s}}+|v_{x}|),~0,~0,~0). \label{eq:boundaryflux}
\ee
We see that all terms in $\hat{\fluxV}_{\rm{HLL}}$ vanish except for the term corresponding to the flux of $x$-momentum -- that is, the component of momentum normal to the face. Thus, the reflective boundary condition (\ref{eq:reflective}) prevents the transport of any quantity besides momentum normal to the face for the HLL Riemann solver; mass, energy and entropy cannot be transported. \lp{This momentum flux corresponds to the contact pressure exerted on the face.} The derivation for the Harten-Lax-van Leer-Contact (HLLC) solver \citep{1994ShWav...4...25T} proceeds in a similar fashion. \lp{\sout{The condition $\flux_{x,\rm{L}}(v_{x}) = \flux_{x,\rm{R}}(v_{x})$ and $\state_{x,\rm{L}}(v_{x}) = \state_{x,\rm{R}}(v_{x})$ corresponds to an inflow boundary condition, although we focus on reflective conditions in this paper.}} The implementation of reflective boundary conditions requires modifications to the Riemann solver and -- for schemes that are accurate to second order or higher in space -- the gradient estimation, both of which are discussed below. \lp{The alternate condition $\flux_{x,\rm{L}}(v_{x}) = \flux_{x,\rm{R}}(v_{x})$ and $\state_{x,\rm{L}}(v_{x}) = \state_{x,\rm{R}}(v_{x})$ corresponds to an inflow boundary condition. However, we focus on reflective conditions in this paper for the reason that inflow or outflow conditions in a moving-mesh code would necessitate the creation or destruction of mesh-generating points, which is not yet implemented in \changaMM. A separate publication (Chang, Davis \& Jiang, submitted to MNRAS) details the development of inflow and outflow conditions for radiation, which is applicable to fluids on a static mesh.}

\subsection{Initialization} \label{sec:init}

In the initialization step at the beginning of the simulation, we declare some user-specified region as the initial boundary. Cells with centres within this region are defined to be a part of the boundary. We will refer to these cells as ``boundary cells'' and to all other cells as ``gas cells.'' The boundary cells will comprise the boundary for the entirety of the simulation, as boundary cells are never added or removed. We define the boundary particles to have equal mass so that the density of the boundary is constant, provided that the density of mesh-generating points is constant within the boundary. The total mass of the boundary is an input parameter in our simulations.

To accurately produce the desired boundary shape, it is necessary to have a sufficiently high density of mesh-generating points near the edges of the boundary. To this end, it is sometimes beneficial to initialize the simulation with a larger resolution in the region containing the boundary than in the rest of the simulation box. We demonstrate this in section \ref{sec:cee}. In principle, a boundary of arbitrary shape can be constructed given a sufficiently high density of mesh-generating points.

\subsection{Edge Cells} \label{sec:edge}

We define all gas cells which share a face with a boundary cell as ``edge cells.'' These cells function as normal gas cells for the most part; however, the motion of their mesh-generating points always matches that of the boundary. The result of this is a layer of gas cells which do not move with respect to the boundary, as in S10. This is important because the surface of the boundary is comprised of the shared faces between the edge cells and boundary cells, which we refer to as ``boundary faces.'' The boundary faces are redrawn with every Voronoi computation, which happens several times during each time-step. Therefore, if the mesh-generating points were allowed to flow past the boundary, the surface of the boundary would deform, which is undesirable. Although the edge cells are the only gas cells which have direct contact with the boundary, they communicate its effects to the rest of the gas through their properties and motion.

\subsection{Gradient Estimation and Linear Reconstruction} \label{sec:gradient}

A second-order accurate code in space demands an appropriate estimate for the state vector at the face centres of each cell. The state vector on the face between the $i$th cell and its $j$th neighbor is taken to be
\be
\tilde{\state}_{ij} = \state_{i} + (\facecentroid_{ij} - \cellcentroid_{i}) \cdot \nabla \state_{i}, \label{eq:facestate}
\ee
where $\facecentroid$ and $\cellcentroid_{i}$ are the centroids of the face and $i$th cell, respectively. Equation (\ref{eq:facestate}) requires an estimate of the gradient of the state vector in the $i$th cell. To calculate the gradients, we follow the procedure of Steinberg et al. (2016), who improved upon the prescription of S10 in using the Gauss-Green theorem to estimate these gradients. We refer the reader to CWQ17 for a more detailed discussion.

We modify the gradient estimation algorithm for edge cells by modifying the properties of their neighboring boundary cells. Consider an edge cell with state vector $\state_{i}$, fluid velocity $\vel_{i}$ and neighbors with states $\state_{j}$, where the normal vectors of the faces shared with its neighbors are $\normal_{ij}$. For each boundary cell neighbor, we first copy the state vector of the edge cell into the boundary cell $(\state_{j} = \state_{i})$. We then modify $\state_{j}$ such that the component of its fluid velocity $\vel_{j}$ normal to the shared face is equal and opposite to $\vel_{i}$. That is,
\be
\vel_{j} = \vel_{i} - 2 (\normal_{ij} \cdot \vel_{i}) \normal_{ij},
\ee
which enforces the condition (\ref{eq:reflective}) for reflective boundary conditions.

We also modify the slope limiter in the presence of a boundary. \changaMM\ uses a slope limiter that reduces numerical oscillations near strong gradients
\be
\langle \nabla \state \rangle_{i}^{S10} = \alpha_{i}^{S10} \langle \nabla \state \rangle_{i},
\ee
where $\alpha_{i}^{S10}$ is defined in S10. As noted by S10, this limiter is not total variation diminishing (TVD), so spurious oscillations can still occur. For this reason, we follow the suggestion by \citet{2011apjs..197...15d} and apply an additional slope limiter
\be
\langle \nabla \state \rangle_{i}^{DM} = \alpha_{i}^{DM} \langle \nabla \state \rangle_{i}^{S10},
\ee
\begin{flushleft}
where
\end{flushleft}
\be
\alpha_{i}^{DM} = \min( 1, \psi'_{ij} ).
\ee
\begin{flushleft}
Here,
\end{flushleft}
\be
\psi'_{ij} = \begin{cases} \max[\theta (\state_{j}-\state_{i})/\Delta\state_{ij}, 1] & |\Delta\state_{ij}| > 0 \\ 1 & \mbox{otherwise} \end{cases}
\ee
and
\be
\Delta\state_{ij} = \langle\nabla\state\rangle_{i} \cdot (\cellcentroid_{j} - \cellcentroid_{i}).
\ee

This limiter is TVD if $\theta < 0.5$, and setting $\theta = 1$ reduces it to the S10 limiter. For all simulations presented in this paper, we use $\theta = 0.49$. However, when performing a reconstruction on a face between an edge cell and a boundary cell we set $\theta = 0$, effectively turning off this additional slope limiter. We find that this marginally improves the conservation properties of our simulations and the stability of the code.

\subsection{Time-Stepping} \label{sec:timestep}

In PC19, we discuss the implementation of individual time-steps into \changaMM, which we summarize here. The basic (universal) time-stepping algorithm for a second-order accurate (in time) integrator can be broken down into 3 stages, the initial time (a), the half time-step (b) and the full time-step (c).

\begin{enumerate}[(a)]
 \item \textit{Initial time $t=0$}: determine Voronoi cells using current positions of mesh-generating points. Estimate gradients and construct half time-step predictions. Zero out the changes to the charges, e.g., $\delta\charge=0$.
 \item \textit{Half time-step $t=0.5\dt$}: drift Voronoi cells to half time-step positions. Construct Voronoi cells at half time-steps. Compute the gradients and perform linear reconstruction to the half time-step cell faces to compute fluxes.  Perform a Riemann solution and incorporate source terms with the {\it full} time-step. Place these changes into $\delta\charge$.
 \item \textit{Full time-step $t=\dt$}: drift Voronoi cells to full time-step positions. Advance charges to be $\charge^{n+1} = \charge^{n} + \delta\charge$. Reset the state to be the new initial time.
\end{enumerate}

We note that the full time-step for any cell involves actions at a full time-step (a,c) and the half time-step (b). For individual time-steps, we assign to each particle a time-step level which we refer to as a ``rung,'' where each rung has a smaller time-step than the rung below it by a factor of two. Therefore, at the half time-step for some rung $i$, full time-steps are executed for all rungs $j > i$. Cells are allowed to change rungs only at their full time-steps. Additionally, we limit the time-step of any cell to be less than $\sqrt{2}$ of the minimum time-step of its neighbors. This ensures that the time-step changes by no more than a factor of 2 (and the rung by no more than 1) over a distance of two cells. This is similar to the time-step smoothing over a SPH kernel used by \citet{2009apj...697l..99s}, and we find that it allows for stable integrations in \changaMM. PC19 finds that individual time-stepping decreases the computation time by a factor of 4 to 5 for a CEE simulation.

When a moving boundary is present, we place the boundary and edge cells on the highest rung (smallest time-step) of all cells present. Besides improving the stability of the code, this mitigates the risk of gas cells penetrating into the boundary or other pathological behaviors. Because the number of edge cells is typically small relative to the total number of cells, we find that this does not substantially slow down the simulation code.

\subsection{Riemann Solver} \label{sec:riemann}

\changaMM\ uses either an HLLC or HLL approximate Riemann solver to estimate the the flux through each face of a cell at the half time-step. We have made several modifications to the Riemann solver in the presence of a boundary.

\begin{enumerate}[(i)]
    \item A Riemann solution is not performed for boundary cells. Because the state vector $\state$ of a boundary cell is determined by those of its neighbors, as described in section \ref{sec:gradient}, there is no need to compute the time rate of change of the boundary cell state vectors. \\
    
    \item For boundary faces, we modify the state vector $\state$ as well as the flux function $\flux$ inside of the boundary cell according to (\ref{eq:reflective}). The reconstruction gives the states on either side of the boundary face in the edge cell and boundary cell $\state_{\rm{e}}$, $\state_{\rm{b}}$ as well as the flux functions $\flux_{\rm{e}}$, $\flux_{\rm{b}}$. We rotate these such that the normal direction to the face lies along the $x$-direction. To enforce reflective boundary conditions, we ignore the $\state_{\rm{b}}$ and $\flux_{\rm{b}}$ given by the reconstruction and instead set them using condition (\ref{eq:reflective}). Denoting the (rotated) $x$-velocity in the edge cell as $v_{x,\rm{e}}$, we require
    \be
    \state_{\rm{b}} = \state_{\rm{e}}(-v_{x,\rm{e}}) \quad\text{and}\quad \flux_{\rm{b}} = \flux_{\rm{e}}(-v_{x,\rm{e}}). \label{eq:trans}
    \ee
    The new $\state_{\rm{b}}$ and $\flux_{\rm{b}}$ along with $\state_{\rm{e}}$ and $\flux_{\rm{e}}$ are then used by the Riemann solver to estimate the flux $\hat{\fluxV}$ across the face, and the result is used in equation (\ref{eq:time evolution}) to evolve the charges in time. \\
    
    \item The contact pressure $\pcont$ at each boundary face, computed by the Riemann solver as in (\ref{eq:boundaryflux}), is given by
    \be
    \pcont = \bar{P} - \frac{1}{2} \normal \cdot (\vel_{\rm{b}}-\vel_{\rm{e}}) \bar{\rho} \bar{c_{\rm{s}}}
    + \begin{cases} 0 & \normal \cdot \vel_{\rm{e}} \leq 0 \\ 2 \bar{\rho} (\normal \cdot \vel_{\rm{e}})^{2} & \normal \cdot \vel_{\rm{e}} > 0 \end{cases}
    \ee
    with the condition that the contact pressure must be non-negative. Here $P$, $\normal$, $\rho$, $c_{\rm{s}}$, are the pressure, normal to the face in the direction of the boundary cell, density and sound speed. The bar denotes an average of the values on each side of the face which are obtained from the state vectors $\state_{\rm{e}}$ and $\state_{\rm{b}}$. For a face with area $A$, this results in a hydrodynamic force
    \be
    \force_{\rm{hy}} = \normal A \pcont \label{eq:force}
    \ee
    at the face. We sum all such forces at boundary faces to find the total hydrodynamic force on the boundary.
\end{enumerate}

\subsection{Gravity Solver} \label{sec:gravity}

\changaMM\ uses a tree-based gravity solver originally implemented in \gasoline\ which uses multipole moments up to fourth order (hexadecapole) to represent the mass distribution within cells at each level of the tree \citep{wadsley+04}. After a gravity solution is performed, we intercept the gravitational accelerations $\accel_{i}$ on all boundary particles and average them to find the gravitational acceleration of the boundary $\accel_{\rm{gr}}$.

For all particles present regardless of type, \changaMM\ uses spline softening to soften the gravitational forces within radius $h$ of the particle. All gas cells in \changaMM\ have equal softening lengths $h_{\rm gas}$, defined at the start of the simulation as the mean separation between cell centres in the region of highest density
\be
h_{\rm gas} = 2(m_{\rm{max}} / \rho_{\rm{max}})^{1/3}.
\ee
Here $\rho_{\rm{max}}$ is the maximum density and $m_{\rm{max}}$ is the cell mass at the location of $\rho_{\rm{max}}$. We require boundary cells to have this same softening length, which guarantees that the self-gravity of the boundary does not impact its motion.

\subsection{Motion of Boundary} \label{sec:motion}

The motion of our boundaries is reactive to both hydrodynamic and gravitational forces. The hydrodynamic force is calculated during the half time-step Riemann solution, and is the sum of all forces on the boundary faces (\ref{eq:force}). For a boundary with mass $m_{\rm{b}}$, this results in an acceleration
\be
\accel_{\rm{hy}} = \frac{1}{m_{\rm{b}}} \sum_{j} \normal_{j} A_{j} P_{{\rm con},j}.
\ee
We combine this with $\accel_{\rm{gr}}$ to get the total acceleration of the boundary
\be
\accel_{\rm{b}} = \accel_{\rm{gr}} + \accel_{\rm{hy}}.
\ee
At the drift step, the boundary velocity
\be
\vel_{\rm{b}} = \accel_{\rm{b}} \dt
\ee
is used to update the positions of all boundary and edge cells. Unlike normal gas cells, the boundary and edge cells do not receive kicks in the direction of the fluid velocity nor the velocity corrections described in CWQ17. \lp{\sout{In contrast to the Euler equations, the equations describing the motion of the boundary are not explicitly momentum-conserving. For this reason, the conservation of the total momentum in our simulation box -- including the boundary momentum $m_{\rm{b}} \vel_{\rm{b}}$ -- is resolution-dependent, as we discuss in section \ref{sec:test}.}}

\section{Test Problems} \label{sec:test}

\subsection{Supersonic Wind} \label{sec:wind}

\begin{figure*}
  \includegraphics[width=1.0\textwidth]{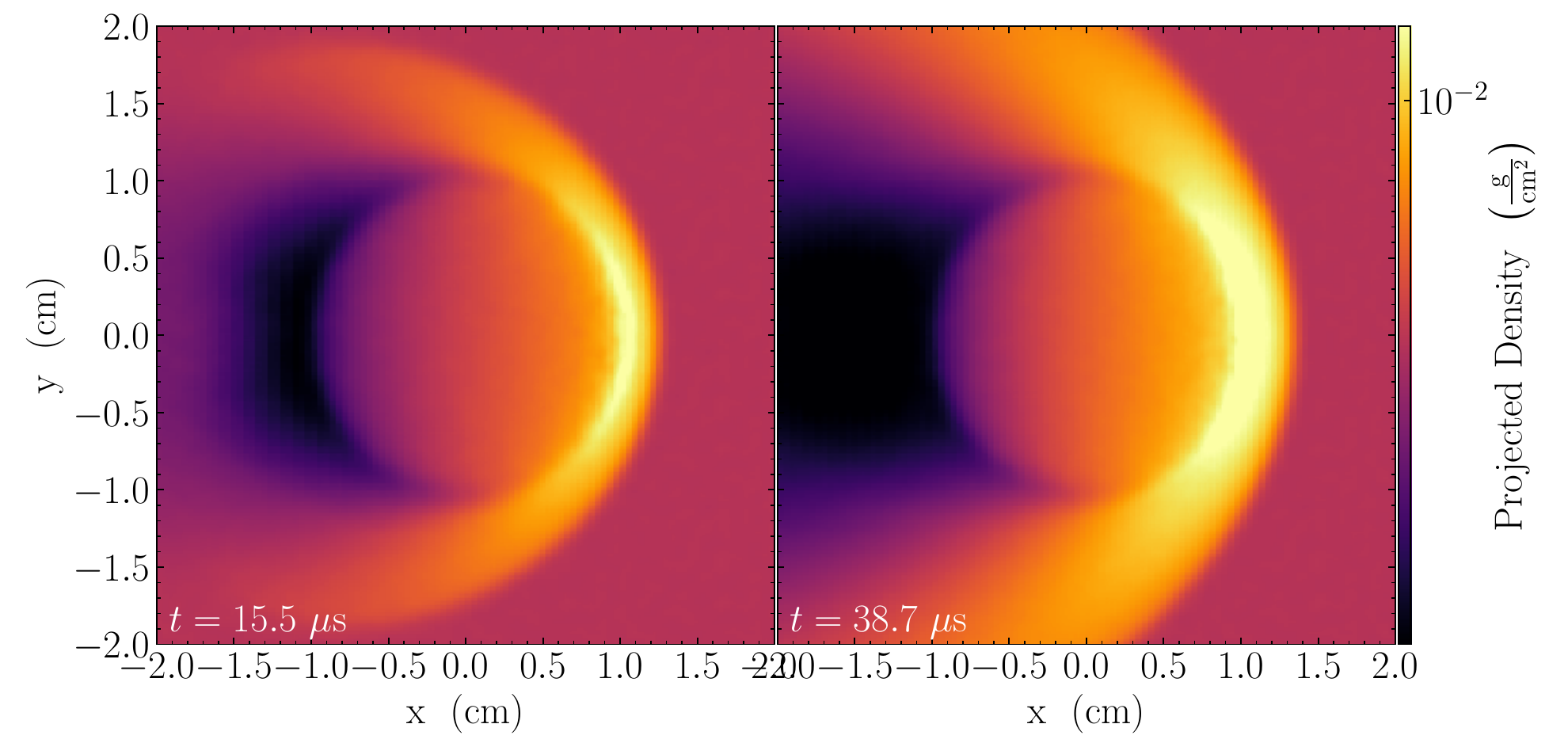}
  \includegraphics[width=1.0\textwidth]{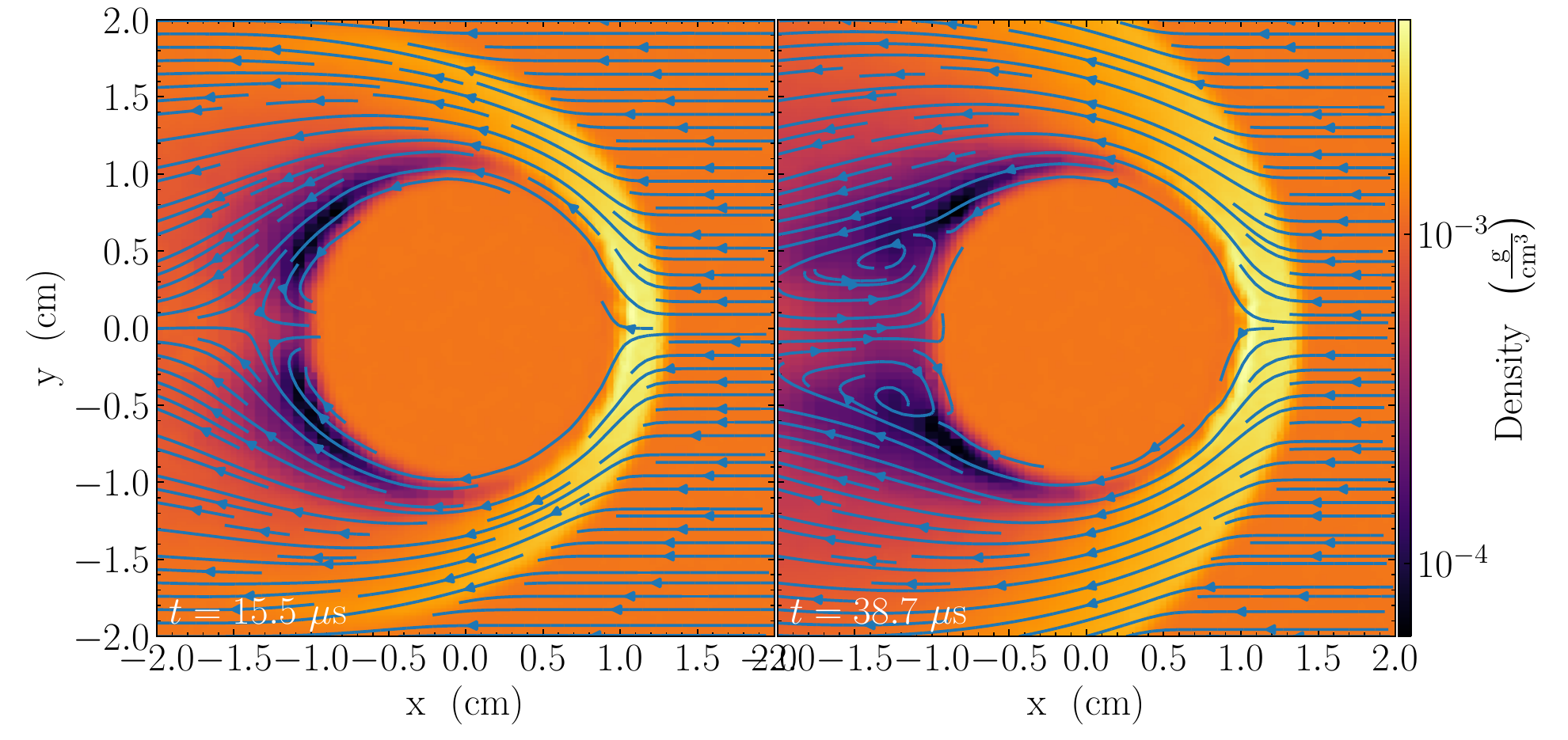}
    \caption{The supersonic wind test case at $t = 15.5~\mu$s \textit{(left)} and $t = 38.7~\mu$s \textit{(right)}. \textit{(Top)} Density projections along the $z$-axis, showing the bow shock as it strengthens. \textit{(Bottom)} Density slices through the $z=0$ plane with streamlines. Here, the stagnation point and vortices are visible.
    \label{fig:wind}}
\end{figure*}

We begin by simulating a sphere moving through air at supersonic speed. The air is initialized with atmospheric conditions $\rho = 0.001225$ g cm$^{-3}$, $T = 288$ K, $\gamma = 1.4$ and mean molecular weight $\mu = 29$ g/mol. Our simulation box has dimensions (10, 6, 6) cm and our spherical boundary has a radius of 1 cm. We use an HLLC Riemann solver for this test and do not include gravity. For all simulations presented in this paper, we use an adiabatic ideal gas equation of state.

We initialize the positions of our mesh-generating points using a pre-computed glass distribution of particles embedded in a 3-D cube. This glass distribution is periodically replicated to produce a sufficient number of particles for our simulation. Low spatial resolution is problematic for this test case as large mesh cells cause the surface of the boundary to be rough and jagged, which can eventually lead to unphysical values for the state vector $\state$. In all test cases, we find that the stability of the code depends largely on the smoothness of the boundary surface. When the boundary surface is well-resolved, we find that our code is robust against unphysical results, even in extreme situations. For this test case, our simulation contains $1.5\times10^{6}$ gas and $1.7\times10^{4}$ boundary cells.

Our boundary moves through the air with Mach number $M=2$, shown in Fig. \ref{fig:wind}. The top left panel shows a density projection at $t = 15.5~\mu$s, where we can see the formation of a bow shock. The bottom left panel shows a density slice through the $z=0$ plane overplotted with the streamlines, where we see turning of the streamlines across the shock as well as a stagnation point behind the boundary. In the panels on the right, we show the same plots at $t = 38.7~\mu$s. Here the shock has strengthened and we see a void in the path cleared by the sphere. In addition, the turbulence behind the boundary has now evolved into vortices. Qualitatively, these are the results that one would expect from this test case. This is encouraging, as the problem of an object moving through a dense gas appears in many astrophysical settings.

\subsection{Sod Shock Tube} \label{sec:sod}

\begin{figure}
  \includegraphics[width=0.5\textwidth]{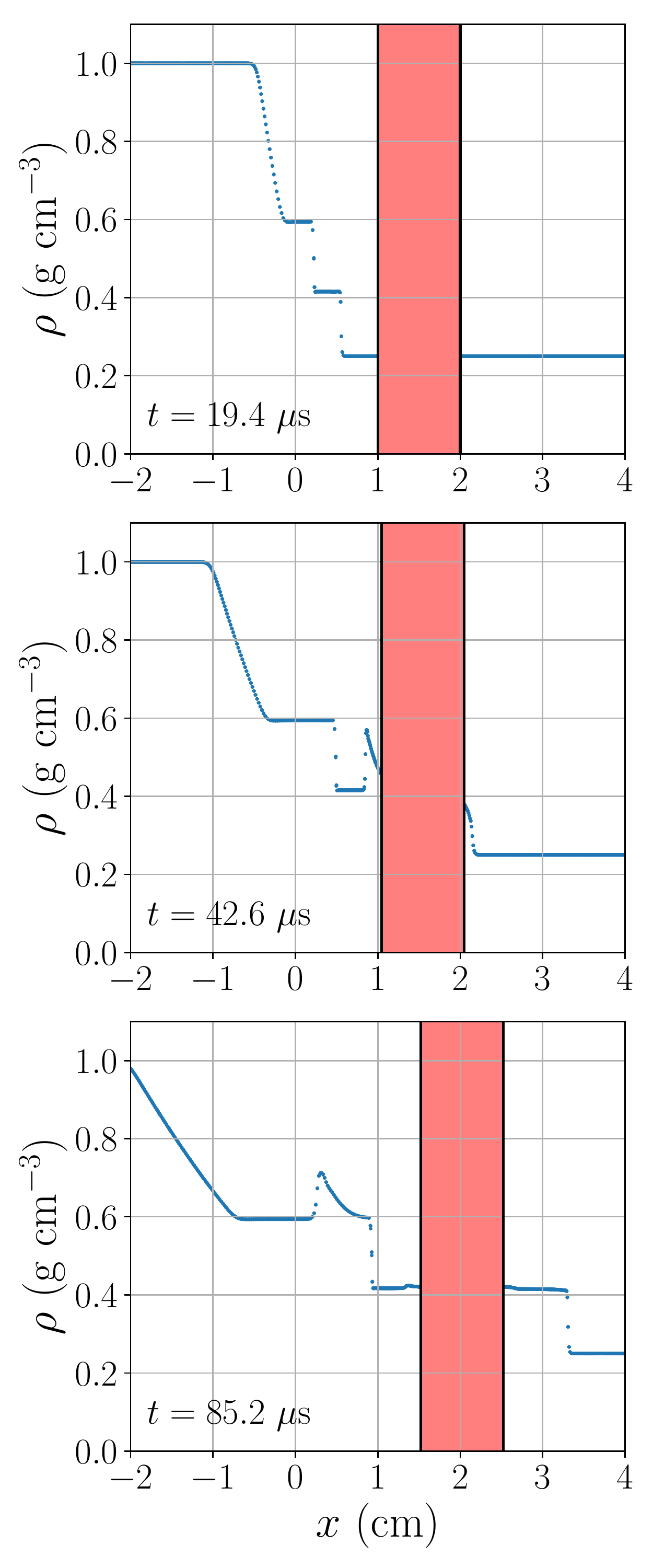}
    \caption{Gas density in our Sod shock tube test at $t = 19.4$, 42.6 and 85.2 $\mu$s. \textit{(Top)} We initially see a shock front propagating to the right, followed by a contact discontinuity, as well as an expanding rarefaction wave. \textit{(Middle)} The shock front then collides with the piston (shaded in red) and reflects back to the left. \textit{(Bottom)} The piston is pushed to the right, creating a transmitted shock; the reflected shock passes the contact discontinuity.
    \label{fig:sod}}
\end{figure}

\begin{figure}
  \includegraphics[width=0.5\textwidth]{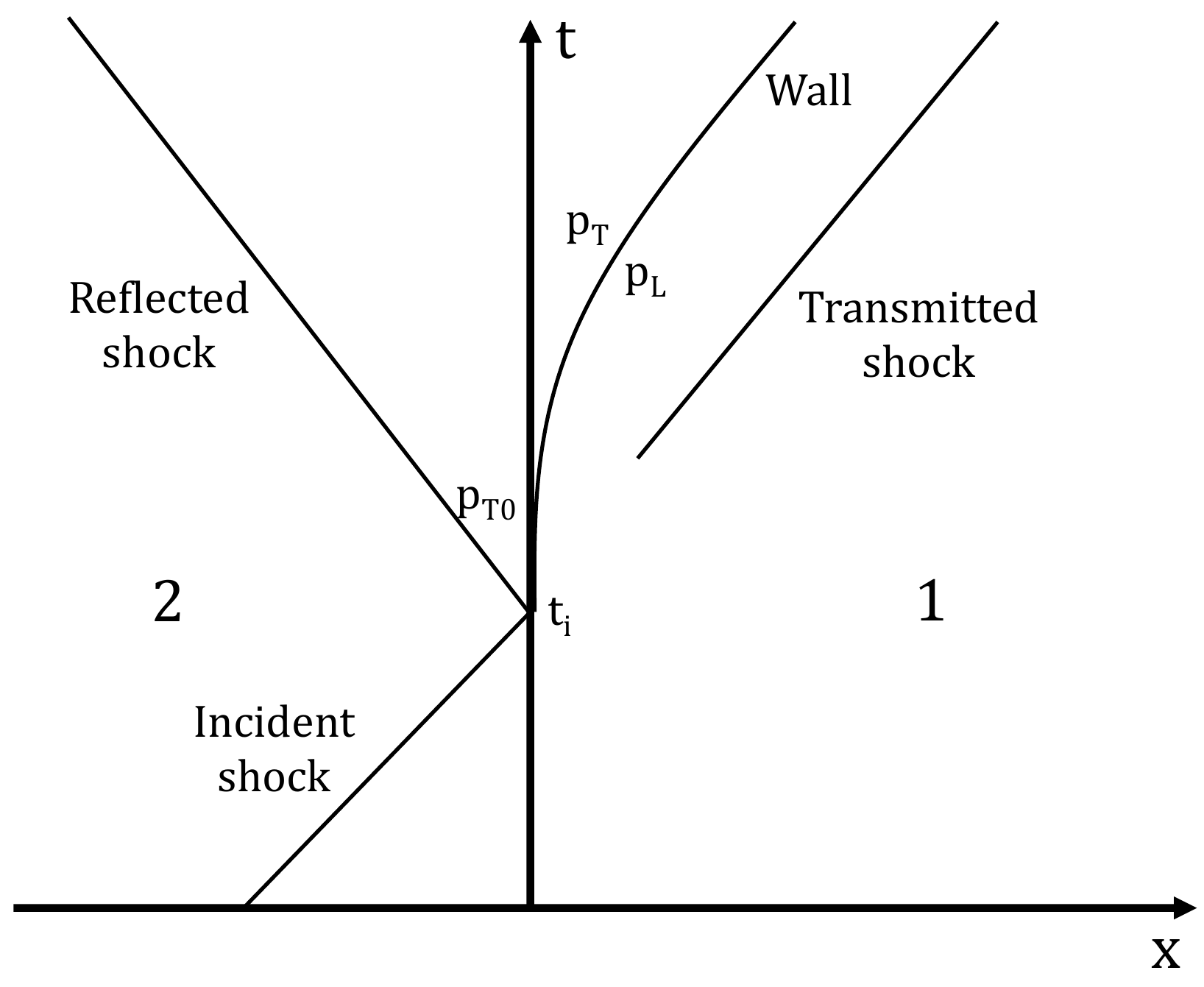}
    \caption{Wave diagram showing the paths of the wall and shock waves. The regions 1 and 2, which denote the states on the right and left sides of the wall before disruption by the transmitted or reflected shocks, are shown as well as the pressures relevant to (\ref{eq:dudt}).
    \label{fig:shockdiagram}}
\end{figure}

We now perform a Sod shock tube test using a simulation box of dimension (16, 1, 1) cm centered at (0, 0, 0). On the left side of the box $(x<0)$, the gas has density $\rho = 1$ g cm$^{-3}$ and temperature $T = 2.5$ K, while the right side $(x\geq0)$ has $\rho = 0.25$ g cm$^{-3}$ and $T = 1.8$ K. The gas is initially stationary $(\vel = 0)$ with $\gamma = 1.4$. We use a periodic simulation box, so the shock propagates through the tube in both directions. These conditions are similar to those used by CWQ17 to validate \changaMM. We choose our moving boundary to be a piston of mass $m = 0.1$ g (1 per cent of the mass of the gas in the tube) that extends from $x=1$ cm to $x=2$ cm and spans the tube in the $y$- and $z$-directions. The piston is free to move but is initially at rest. We use a rectangular mesh -- an appropriate choice because of our rectangular piston -- with 160 cells in the $x$-direction and 10 cells in the $y$- and $z$-directions, for a total of 16,000 cells. \lp{\sout{We also perform four higher-resolution runs with 320, 640, 1280 and 2560 cells in the $x$-direction while still using 10 cells in the $y$- and $z$-directions. To keep the cells cubical at higher resolutions, we reduce the width of the tube in the $y$- and $z$-directions appropriately. We also adjust the mass of the piston such that the ratio of its mass to that of the gas is preserved.}}

In Fig. \ref{fig:sod}, we show the density as a function of position at $t = 19.4$, 42.6 and 85.2 $\mu$s. We initially see the formation of the shock front, contact discontinuity, and rarefaction wave. At $t=34.8$ $\mu$s, the shock front impacts the piston (shaded in red) and reflects back to the left. This impact also pushes the piston to the right, inducing a transmitted shock.

\begin{figure}
  \includegraphics[width=0.5\textwidth]{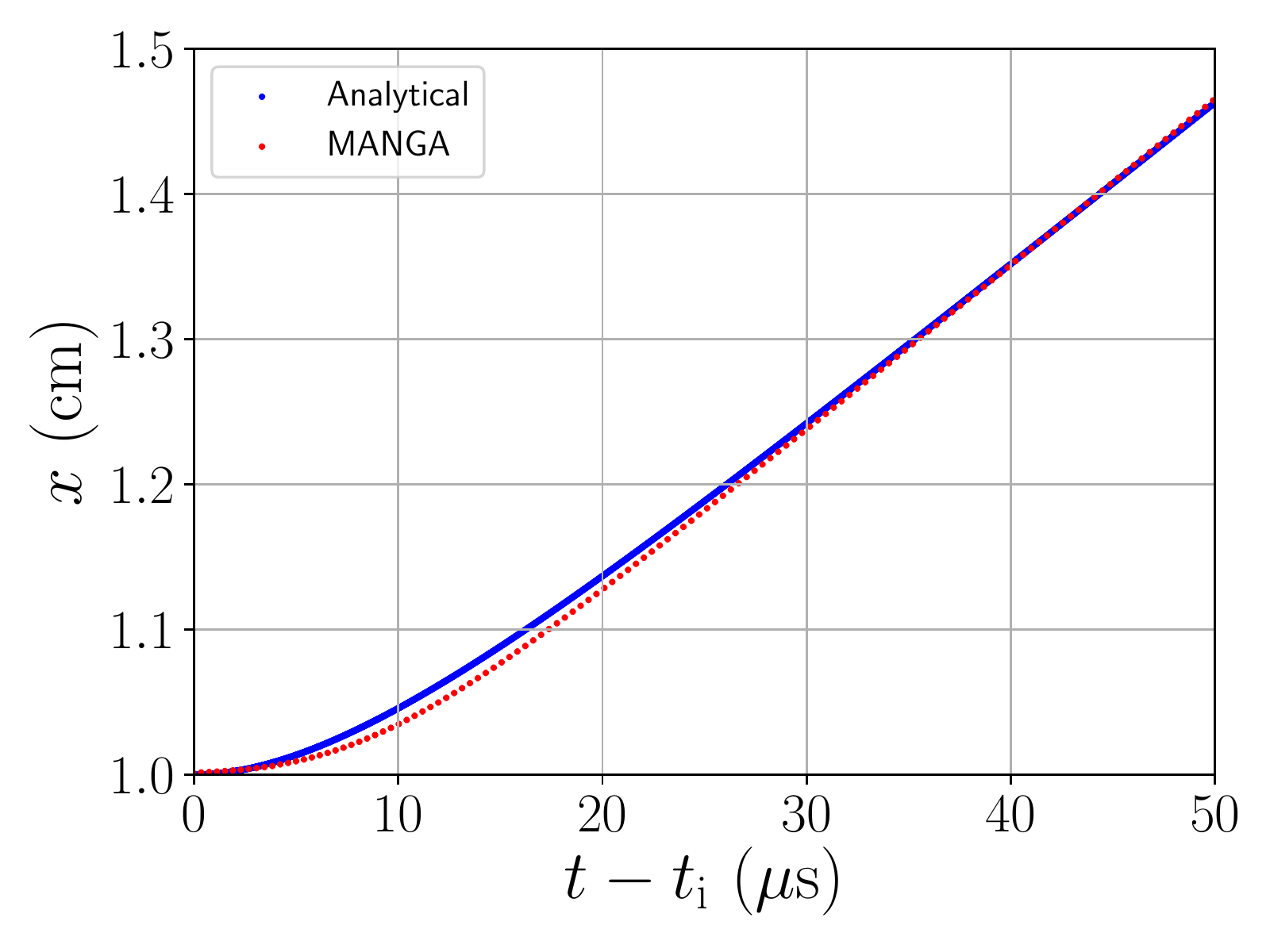}
    \caption{An analytical solution for the response of a movable wall to a shock wave (blue) following \citet{1957JFM.....3..309M} compared with the results from our Sod shock tube test case in \changaMM\ (red). On the horizontal axis is the time after shock impact $t_{\rm i}$.
    \label{fig:pistontrack}}
\end{figure}

\lp{We validate the motion of the piston using an analytical result for the impact of a shock wave on a movable wall. Following \citet{1957JFM.....3..309M}, the motion of the wall is described by}
\be
\frac{du_{\rm w}}{dt} = \frac{p_{\rm T}-p_{\rm L}}{p_{\rm T 0}-p_{1}} \left( \frac{du_{\rm w}}{dt} \right)_{t=t_{\rm i}}. \label{eq:dudt}
\ee
\lp{Here $u_{\rm w}$ is the wall velocity, $p$ is the gas pressure and $t_{\rm i}$ is the time of shock impact. The subscripts L and T correspond to the leading and trailing edges of the wall (Fig. \ref{fig:shockdiagram}). The regions 1 and 2 are the states on the right and left sides of the wall before disruption by the transmitted or reflected shocks. Furthermore, $p_{\rm T 0}$ is the pressure on the trailing edge just after shock reflection. We use the isentropic relations as well as the Riemann invariants}
\be
P = \frac{2}{\gamma-1}c_{\rm s}+v \quad\text{and}\quad Q = \frac{2}{\gamma-1}c_{\rm s}-v
\ee
\lp{to obtain the relations}
\be
p_{\rm L} = p_{1} \left(\frac{Q_{1}+u_{\rm w}}{Q_{1}}\right)^{2\gamma / (\gamma-1)}
\ee
\lp{and}
\be
p_{\rm T} = p_{\rm T 0} \left(\frac{P_{2}-u_{\rm w}}{P_{2}}\right)^{2\gamma / (\gamma-1)}.
\ee
\lp{We can express the initial wall acceleration pressure as}
\be
\left( \frac{du_{\rm w}}{dt} \right)_{t=t_{\rm i}} = (p_{\rm T 0}-p_{1}) \frac{A}{m},
\ee
\lp{where $A$ is the cross-sectional area of the wall. Putting this all into (\ref{eq:dudt}), we arrive at}
\be
\frac{du_{\rm w}}{dt} = \frac{A}{m} \left[ p_{\rm T 0} \left(\frac{P_{2}-u_{\rm w}}{P_{2}}\right)^{2\gamma / (\gamma-1)} - p_{1} \left(\frac{Q_{1}+u_{\rm w}}{Q_{1}}\right)^{2\gamma / (\gamma-1)} \right], \label{eq:dudt2}
\ee
\lp{which can be numerically integrated twice with the boundary conditions $u_{w}=0$ and $x=1$ cm at $t=t_{\rm i}$ to find the path of the wall. We obtain the relevant pressures and Riemann invariants for (\ref{eq:dudt2}) from our simulation output. The results are shown in Fig. \ref{fig:pistontrack} alongside those obtained in \changaMM, showing that the two are consistent.}

\lp{\sout{Because the system is initially at rest, it has initial momentum $p_{\rm{tot}}=0$. However, the impact of the shock against the boundary can cause the total momentum to depart significantly from zero. We define a momentum error $\pdiff$ as the fractional difference between the gas momentum $p_{\rm{gas}}$ and boundary momentum $p_{\rm{b}}$. We find that $\pdiff$ does not change significantly after the shock impact. Table \ref{tab:sod} shows $\pdiff$ just after shock impact (when the left edge of the piston reaches $x=1.1$ cm) as well as the cell size $\Delta x$ for each of the five runs in the resolution test. A least-squares fit of the data in Table \ref{tab:sod} gives $\pdiff \propto (\Delta x)^{0.98}$, showing that the momentum conservation converges as the spatial resolution.}}



\subsection{Sedov-Taylor Blast Wave} \label{sec:sedov}

In this test problem, we use the same setup as in the Sod shock tube test. However, we now initialize the gas with density $\rho = 1$ g cm$^{-3}$ and temperature $T = 1$ K except for the region $|x| < 0.1$ cm, where we set $T = 1000$ K. This results in a point explosion that is described by the Sedov-Taylor solution. It is known that the Sedov-Taylor solution is self-similar in 3-D, and we briefly show that this also holds in 1-D. Following \citet{1950RSPSA.201..159T}, the appropriate similarity assumptions for an expanding blast wave of constant total energy in 1-D are
\be
p/p_{0} = R^{-1} f_{1}, \quad \rho / \rho_{0} = \psi \quad\text{and}\quad u = R^{-1/2} \phi_{1}.
\ee
Here $R$ is the radius of the blast wave, $u$ is the fluid velocity, $p_{0}$ and $\rho_{0}$ are the ambient pressure and density and $f_{1}$, $\psi$, and $\phi_{1}$ are functions only of similarity variable $\eta = x/R$. In the absence of gravity, we can now cast the Euler equations (\ref{eq:continuity}), (\ref{eq:momentum}) and (\ref{eq:energy}) into the one-dimensional forms
\be
\ddt{u} + u \ddx{u} = -\frac{p_{0}}{\rho} \ddx{(R^{-1} f_{1})}, \label{eq:eom}
\ee
\be
\ddt{\rho} + u \ddx{\rho} + \rho \ddx{u} = 0 \label{eq:cont}
\ee
\begin{flushleft}
and
\end{flushleft}
\be
(\ddt{} + u\ddx{})(p \rho^{-\gamma}) = 0. \label{eq:eos}
\ee
These equations are satisfied if $\frac{dR}{dt} = A R^{-1/2}$, where $A$ is a constant. If we additionally define $a^{2} = \gamma p_{0} / \rho_{0}$, $f = f_{1} a^{2}/A^{2}$ and $\phi = \phi_{1}/A$, then (\ref{eq:eom}), (\ref{eq:cont}) and (\ref{eq:eos}) become
\be
\frac{1}{2}\phi - \eta\phi' + \phi\phi' = \frac{1}{\gamma\psi}(f - f'), \label{eq:eom2}
\ee
\be
-\eta\psi' + \phi\psi' + \phi'\psi = 0 \label{eq:cont2}
\ee
\begin{flushleft}
and
\end{flushleft}
\be
-\gamma\eta f \psi' + f \psi^{-\gamma} + \eta f' \psi^{-\gamma} + \gamma f \phi \psi' - f' \phi\psi^{-\gamma} = 0, \label{eq:eos2}
\ee
respectively. Here a prime denotes a derivative with respect to $\eta$. We see that (\ref{eq:eom2}), (\ref{eq:cont2}) and (\ref{eq:eos2}) are independent of $R$; they depend only on similarity variable $\eta$ and the gas properties $p_{0}$, $\rho_{0}$ and $\gamma$. Therefore, this is a self-similar solution. Furthermore, because (\ref{eq:eom2}), (\ref{eq:cont2}) and (\ref{eq:eos2}) are all first-order, only one boundary condition is needed for each of $f$, $\phi$ and $\psi$ to find a solution. These are given by the Rankine-Hugoniot shock jump conditions, which for a sufficiently strong shock take on the asymptotic forms
\be
\psi = \frac{\gamma+1}{\gamma-1}, \quad f = \frac{2\gamma}{\gamma+1} \quad\text{and}\quad \phi = \frac{2}{\gamma+1}
\ee
at the shock front ($\eta = 1$).







\begin{figure}
  \includegraphics[width=0.5\textwidth]{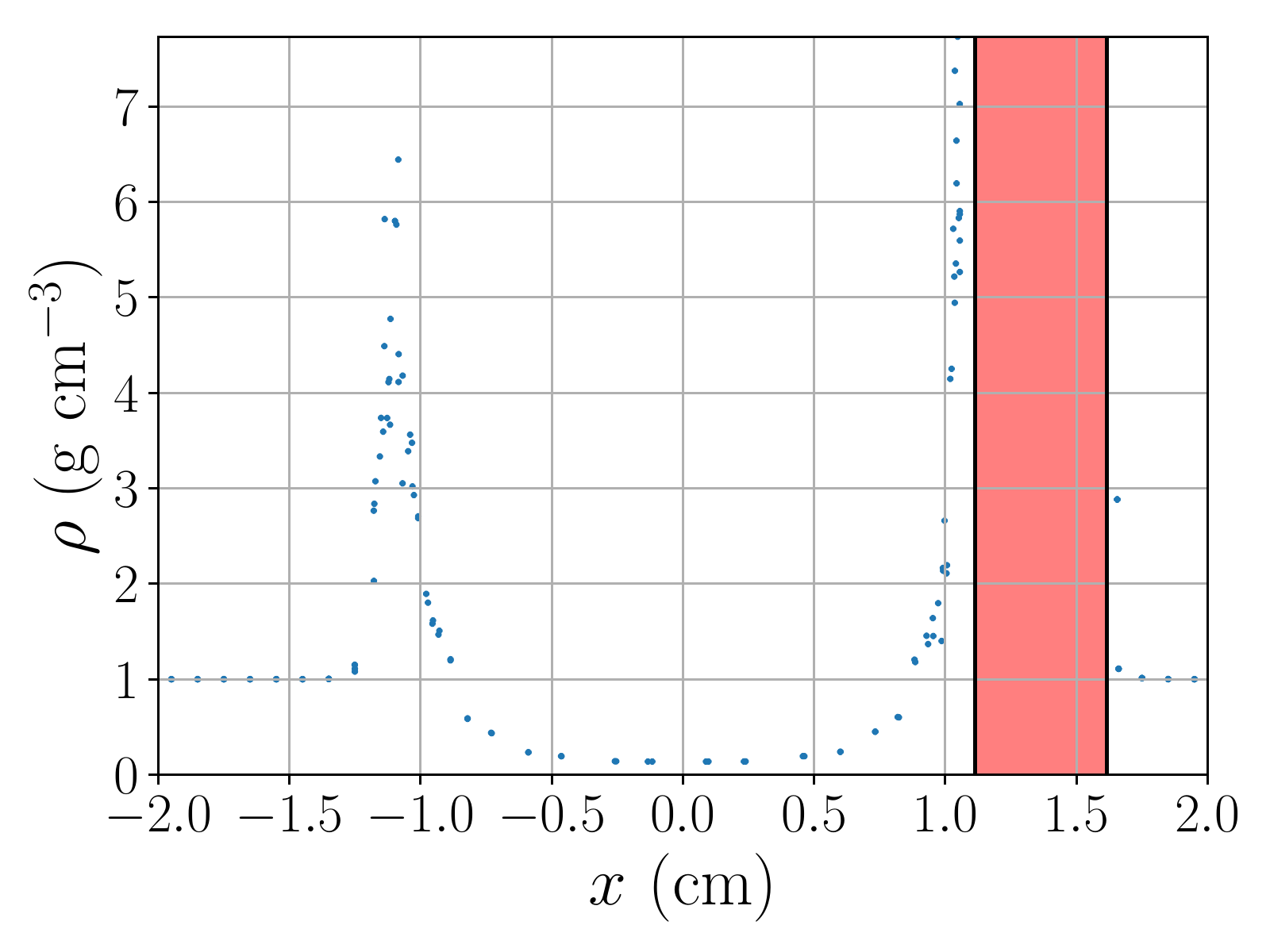}
    \caption{Gas density in the Sedov-Taylor blast wave test at $t = 6$ $\mu$s, restricted to particles with $|y| < 0.1$ and $|z| < 0.1$ cm to reduce the particle noise. The blast wave is colliding with a piston (shaded in red) and pushing it to the right. The gap between the left side of the piston and the mesh-generating points comprising the shock front is due to the edge cell centres maintaining their original distance from the boundary.
    \label{fig:sedov}}
\end{figure}

We again choose a piston as our moving boundary, extending from $x=1$ to $x=1.5$ cm with mass $m = 0.5$ g. We plot the density as a function of position in Fig. \ref{fig:sedov} at $t = 6$ $\mu$s, showing only particles with $|y|<0.1$ and $|z|<0.1$ cm in order to reduce the particle noise. The blast wave quickly forms and propagates outward until it impacts the piston, which is shaded in red, pushing it to the right and creating a transmitted shock. A gap is visible between the left side of the piston and the mesh-generating points comprising the shock front, due to the edge cell centres maintaining their original distance from the boundary. \lp{\sout{We again find the that shock impact introduces a substantial momentum non-conservation; using (\ref{eq:pdiff}) as a metric yields $\pdiff = 1.16$.}} \lp{To analyze the conservation properties of our simulation, we consider the fractional error in the conservation of momentum of the system $\pdiff$, including the contributions from both the gas and the piston. Here, we find $\pdiff = 4.68\times10^{-10}$.} However, we also see from the spacing of the mesh-generating points that the spatial resolution of the blast wave is poor, and thus conduct a resolution test to determine its effect on $\pdiff$. Rather than changing the cell size\lp{\sout{and tube dimensions as with the Sod shock tube}}, we exploit the self-similarity of the Sedov-Taylor solution. We perform two additional runs in which the piston is initially placed further to the right so that the number of cells $N_{\rm c}$ resolving the blast wave is greater. The mass and width of the piston are taken to be proportional to $R$ to preserve the self-similarity of the system. Specifically, we use pistons of mass $m=1$ and $m=1.5$ g, extending from $x=2$ to $x=3$ and from $x=3$ to $x=4.5$ cm, respectively. These are shown in Table \ref{tab:sedov} along with $\pdiff$, which we calculate when the left edge of the piston has moved from its initial position by 10 per cent. \lp{We find that higher resolution improves momentum conservation as $\pdiff \propto N_{\rm c}^{-1.7}$.}

\begin{table}
\caption{Piston initial position, piston mass $m$, number of cells resolving the blast wave $N_{\rm c}$ and momentum error $\pdiff$ in our Sedov-Taylor blast wave resolution test. We calculate $\pdiff$ when the left edge of the piston has moved from its initial position by 10 per cent.}
\label{tab:sedov}
\begin{center}
\begin{tabular}{c c c c}
\hline
\hfill $x$ Position (cm) & $m$ (g) & $N_{\rm c}$ & $\pdiff$ \\
\hline
\hfill 1 -- 1.5 & 0.5 & 10 & $4.68\times10^{-10}$ \\
\hfill 2 -- 3 & 1 & 20 & $1.27\times10^{-10}$ \\
\hfill 3 -- 4.5 & 1.5 & 30 & $7.44\times10^{-11}$ \\
\hline
\end{tabular}
\end{center}
\end{table}

\lp{\sout{Because we can define a length scale $L$ associated with the blast wave, it is possible to draw conclusions about the spatial resolution needed to enforce momentum conservation. Let us define $L$ as the distance from the shock front to the point behind the shock where the density is reduced by one $e$-folding, and $N_{\rm{c}} = L / \Delta x$ as the number of cells (in the $x$-direction) resolving this region. A numerical solution of a 1-D Sedov-Taylor blast wave by \citet{kamm_sedov} for a gas with $\gamma=1.4$ yields $L = 0.16 R$. Using this and our data for $\pdiff$ in Table \ref{tab:sedov}, we obtain the relation $\pdiff = 0.259 L^{-0.815}$, which we rearrange to arrive at}}

\lp{\sout{Thus, for a desired accuracy $\pdiff$ we have a condition on the minimum number of cells necessary to resolve the blast wave.}}

\subsection{Common Envelope Evolution} \label{sec:cee}

\begin{figure*}
  \includegraphics[width=1.0\textwidth]{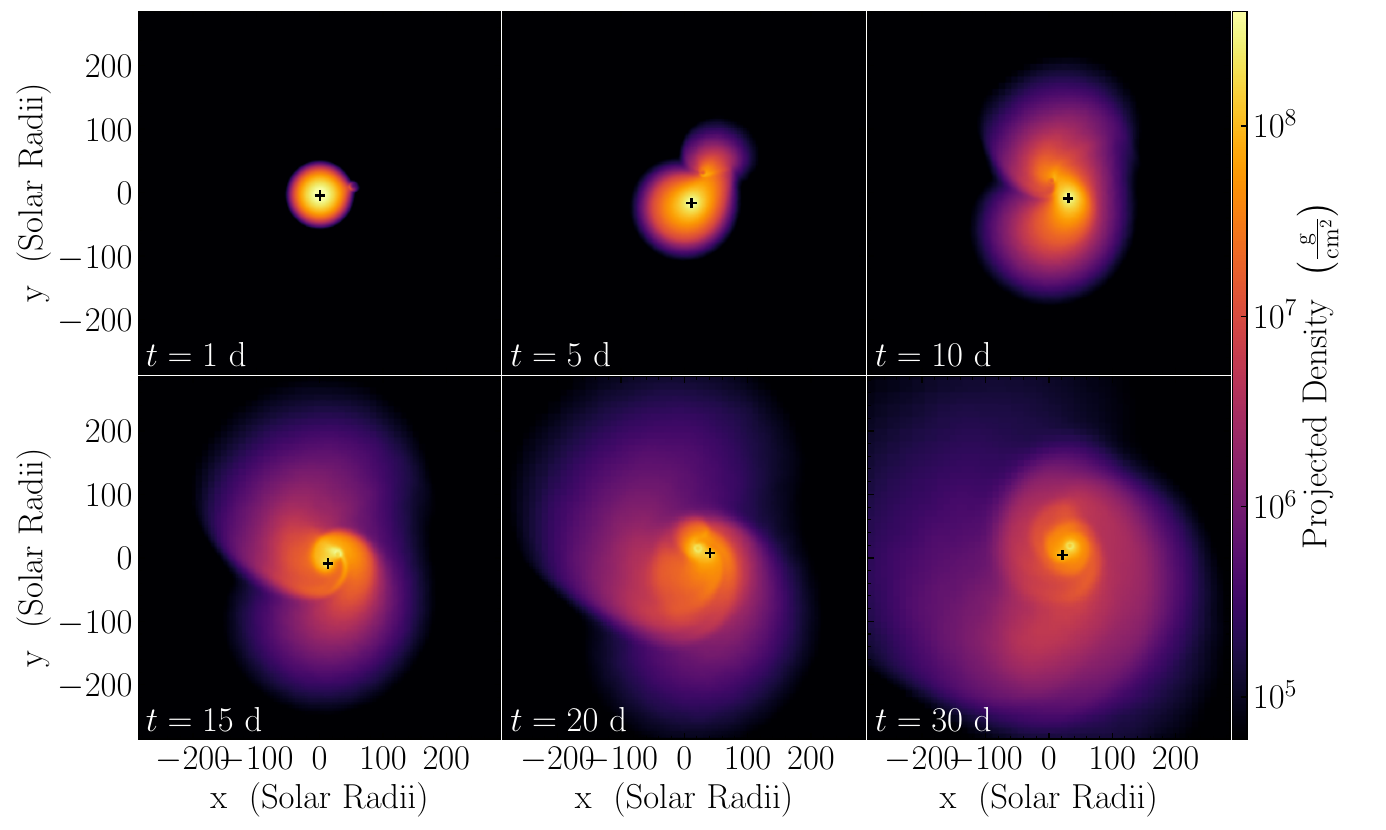}
    \caption{Density projections along the $z$-axis in our CEE simulation. The $+$ sign marks the red giant core.
    \label{fig:frames}}
\end{figure*}

In this test problem we demonstrate the applicability of moving boundaries to astrophysical systems by simulating common envelope evolution (CEE), a brief phase in the life of a binary star system. In CEE, two stars -- a giant star and a much smaller companion -- share a common gaseous envelope \citep[for a review see][]{2013A&ARv..21...59I}. As the companion and the core of the giant orbit each other, loss of orbital energy and angular momentum drive the two stellar centres close to one another and eject the envelope from the system. CEE is a critical process in the lives of these binary systems and is responsible for progenitors of Type Ia supernova (potentially), X-ray binaries, double white dwarfs, double neutron stars and possibly the merging double black hole and double neutron star systems discovered by Advanced LIGO \citep{2013A&ARv..21...59I,2016Natur.534..512B}.

In particular, multiple groups have run 3-D numerical simulations of CEE. Several numerical methods have been tried, including smooth particle hydrodynamics \citep{2015MNRAS.450L..39N,2012ApJ...744...52P}, Eulerian grid codes \citep{2012ApJ...744...52P,2012ApJ...746...74R,2019MNRAS.486.1070C} and moving-mesh codes \citep[PC19]{2016ApJ...816L...9O}. In all of these simulations, the radius of the companion is considered negligibly small compared to the size of the giant, and thus the companion is modeled by a point particle that interacts only through gravity. This approximation is accurate when the companion in question is a compact object such as a black hole or neutron star, however in many systems of interest the companion is a small star with size of order 1 $\rsun$. In such cases, the drag force resulting from the companion encountering the envelope may significantly alter the dynamics of the spiral-in. This is where moving boundaries come into play, as the companion can be modeled as a spherical boundary. \citet{2019arXiv190806195C} and \citet{2019arXiv191013333D} investigated the drag force on the companion in CEE simulations using adaptive mesh refinement codes with point particle companions, and in future work we hope to make comparisons to their results.

The application of \changaMM\ to CEE was the main focus of PC19, and here we use the same binary system. We summarize the simulation setup here, but refer the reader to PC19 for a more detailed discussion. We first use \mesa\ \citep{2011ApJS..192....3P,2013ApJS..208....4P,2015ApJS..220...15P,2018ApJS..234...34P,2019ApJS..243...10P} to evolve a 2 \msun\ star with metallicity $Z=0.02$ from the pre-main sequence to the red giant phase.  We stop when the stellar radius reaches 52 \rsun\ with a helium core of mass $M_{\rm{c}} = 0.36~\msun$. Because of the great difference in density between the helium core and the hydrogen envelope, we model the core as a dark matter particle with gravitational softening length $R_{\rm{c}} = 1.99~\rsun$.

We construct an appropriate particle mesh for the star using the same glass distribution as in section \ref{sec:wind}. We assume that each particle is of approximately equal volume and re-scale them to the appropriate mass based on the computed $M(r)$ from \mesa. The total number of particles representing the star is $8.5\times10^{4}$.  Outside of the star, we include a low density atmosphere of $10^{-13}$ g cm$^{-3}$ with temperature $10^{5}$ K that extends out to the total box size of $3.5\times 10^{14}$ cm ($5000~\rsun$), with periodic boundary conditions at its edges. The total number of particles within the simulation box is $5.4\times10^{5}$.

We place the 1 \msun\ companion in an initially circular Keplerian orbit at the red giant radius $a=52~\rsun$. We assume that, prior to the onset of CEE, tidal forces from the companion have spun up the rotation of the giant \citep{1996ApJ...460L..53S}. Thus, we set the rotation of the giant such that it is tidally locked to the binary orbit.

To create the moving boundary, we assume that the companion is a sphere of radius $4~\rsun$, which is unrealistically large. We use such a large companion as a proof of concept for the methodology of moving boundaries. As discussed in section \ref{sec:init}, it is important to have a sufficiently high density of mesh-generating points near the boundary to accurately produce the desired boundary shape. To this end, we refine our mesh near the initial position of our boundary. Note that this needs only be done during the initialization; because the boundary cells and edge cells move identically, a boundary that is highly resolved initially will remain so indefinitely. We perform our refinement within a cube of side length 10 $\rsun$ centered on the companion. We use the same glass distribution, but scaled such that the cube contains 1000 mesh-generating points. Our boundary comprises 270 of these points, and the rest of the particles in the cube are given the same properties as the external atmosphere. This somewhat cuts into the red giant envelope, however the outer layers of the envelope are very rare and the total mass of the giant is not significantly altered.

\begin{figure}
  \includegraphics[width=0.5\textwidth]{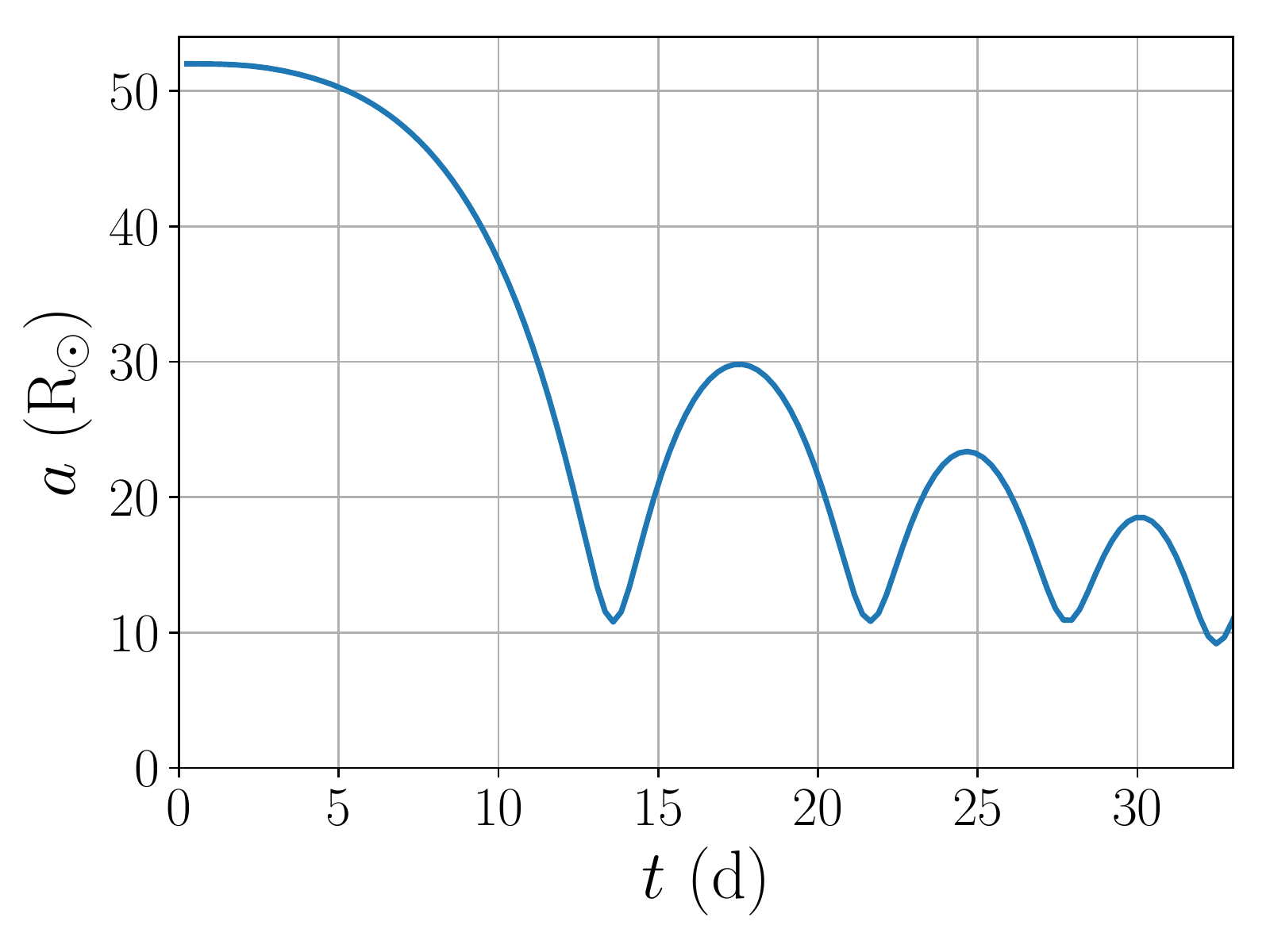}
    \caption{Separation $a$ between the stellar core and companion. The companion quickly plunges into the envelope, reducing the separation between the core and companion by a factor of 5. The orbit then continues to shrink at a slower rate.
    \label{fig:separation}}
\end{figure}

We simulate the binary for 33 d and show several density projections in Fig. \ref{fig:frames} at $t = 1$, 5, 10, 15, 20 and 30 d. For the first 14 days, the companion plunges into the envelope, reducing the separation between the core and companion by a factor of 5 (Fig. \ref{fig:separation}). We see tidal tails of gas ejected from the system by both the core and companion during this plunge. The orbit then continues to shrink for the remainder of the simulation period as orbital energy is transferred to the gas. Spiral shocks facilitating this transfer can be seen in the lower panels of Fig. \ref{fig:frames}. The simulation ends well before the outflow reaches the edge of the simulation box.

\begin{figure}
  \includegraphics[width=0.5\textwidth]{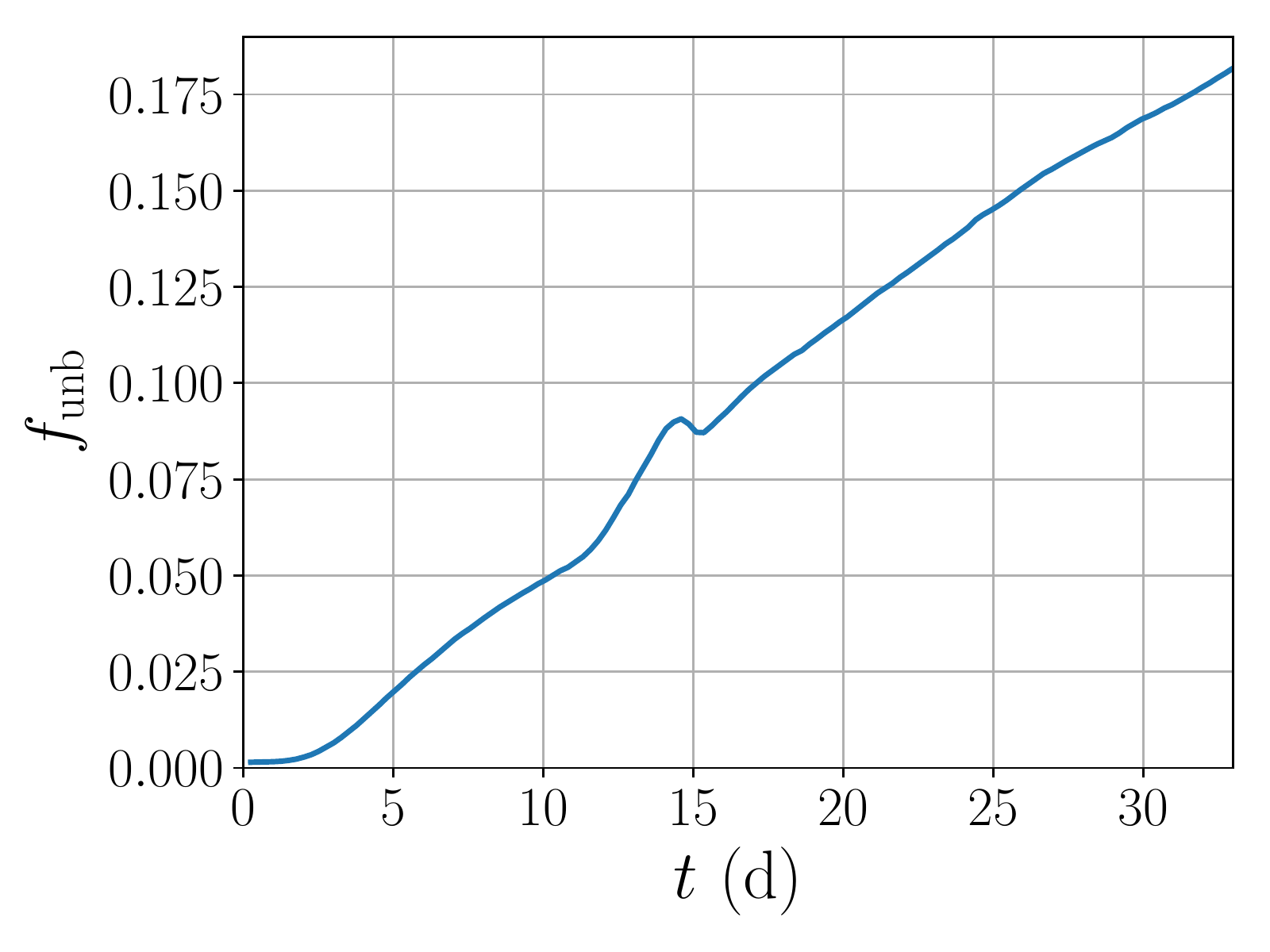}
    \caption{The fraction of mass of the envelope that has acquired enough energy to be unbound from the system. It shows fairly linear behavior over most of the simulation period, with the exception of the first periastron passage at $t \approx 14$ d.
    \label{fig:unbound}}
\end{figure}

The total energy of each gas particle is given by
\be
\Etoti = m_{i} [ \frac{1}{2} (\vel_{i} - \velCM)^{2} + \phi_{i} + I_{\rm{e},i} ] \label{eq:Etot}
\ee
\citep{2015MNRAS.450L..39N}, where $m_{i}$, $\vel_{i}$, $\velCM$, $\phi_{i}$ and $I_{e,i}$ are the mass, velocity, centre of mass (CM) velocity of the bound material, gravitational potential and specific internal energy of each particle. Gas particles with a negative total energy are bound to the binary, while those with a positive total energy are unbound. The kinetic energy is computed relative to the velocity of the CM of the \textit{bound} matter; that is, the bound gas as well as the red giant core. However, because the total energy is needed in order to determine which gas particles are bound, we use an iterative scheme to find the velocity of the CM which is described in PC19. With the total energies of all particles known, we can find the mass of the unbound gas $\munbound$ as a fraction of the total mass of the envelope $m_{\rm env}$,
\be
\funbound = \frac{ \munbound }{ m_{\rm env} }.
\ee
We refer to this as the ejection efficiency, shown in Fig. \ref{fig:unbound}. The ejection efficiency has reached nearly 20 per cent by the end of the simulation period, but is still steadily increasing. The first periastron passage at $t \approx 14$ d temporarily causes an increase in $\funbound$.

This simulation exhibits features similar to those shown in the previous work listed above, but with a lower resolution and a much shorter simulation period. Therefore, the numerical results presented here should only be considered as a validation of our methodology. Nevertheless, this test case shows that using moving and reactive boundaries in CEE simulations is feasible and produces realistic results. We leave a detailed study for future work.

\section{Discussion and Future Directions}\label{sec:discussion}

We describe the implementation of moving and reactive boundary conditions into the moving-mesh code \changaMM. ALE schemes are ideal for moving boundaries because the mesh cells are designed to move with respect to one another; S10 demonstrates this by carrying out a simple 2-D simulation of a curved boundary that moves through a fluid along a fixed path. We describe the hydrodynamics of boundary conditions, which can be realized by modification of the flux function $\flux$ and state vector $\state$ at the boundary surface in both the Riemann solver and gradient estimation. To realize reflective boundary conditions, the state vector and flux function on either side of a boundary face must be equal except for the fluid velocities normal to the face, which must be equal and opposite. For inflow conditions, the fluid velocities are simply equal. We show that all components of the HLL flux across a boundary face are zero except for the momentum normal to the face, meaning that no other conserved variable can be transported across the boundary. We require that the gas cells adjacent to the boundary do not move with respect to the boundary, so that the surface of the boundary does not deform. However, the motion of our boundaries is reactive to external forces. We describe the modifications to our Riemann and gravity solvers to compute the aerodynamic drag and gravitational force on the boundary.

We test our methodology in four cases. The first is a supersonic spherical boundary with Mach number $M=2$ moving through air, forming a bow shock on the leading edge. We see the expected turbulence pattern behind the sphere: a stagnation point appears before evolving into vortices. The next two are the Sod shock tube test and a Sedov-Taylor blast wave, where in both cases the shock front collides with a movable piston. \lp{\sout{Because the equations describing the motion of the boundary are not explicitly momentum-conserving, we find a degree of momentum non-conservation $\pdiff$ in these cases. However, this non-conservation is dependent on the spatial resolution, and for the Sod shock tube we show that it has the convergence property $\pdiff \propto \Delta x$.}} \lp{We validate the motion of the piston in the Sod shock tube case by comparison with the analytical results of \citet{1957JFM.....3..309M}.} For the case of a blast wave we show that the Sedov-Taylor solution is self-similar in one dimension and exploit this fact to find a relationship between the number of cells used to resolve the blast wave and the fractional momentum error. Finally, we demonstrate an astrophysical application of our methodology by simulating CEE in a binary system. All previous work on numerical simulations of CEE have modeled the companion object as a point particle, and moving boundaries allow us to model it as an object of finite size. Though this test simulation has low spatial resolution, it shows the expected behavior of a common envelope phase.

We find that the stability of the code depends largely on the smoothness of the boundary surface. This typically necessitates a high density of mesh-generating points in the vicinity of the boundary at the start of the simulation. In our CEE simulation, we refine the mesh in a cubical region surrounding the initial position of our boundary. Because the number of cells in this cube after the refinement is small compared to the total number of cells, this does not significantly slow our code.

Moving boundaries have yet to find success in astrophysics due to the difficulty of implementation in most codes as well as the relative lack of applications as compared to other fields. However, there are many cases in which they prove useful, \lp{including the study of stellar pulsations as well as any astrophysical object moving through a dense gas}. The first and most straightforward use for this methodology will be to conduct in-depth simulations of CEE using a moving boundary. High-resolution simulations of CEE using the \mesa\ equation of state could shed light on the effects of companion size and inform future work. In particular, the drag force on the companion is expected to depend upon the size of the companion, and could alter the dynamics of the spiral-in phase \citep{2019arXiv190806195C,2019arXiv191013333D}.

\lp{Furthermore, if the creation and destruction of mesh-generating points is implemented into \changaMM, additional applications will become possible.} Because our reflective boundary conditions can be transformed into inflow conditions with a few sign changes in the code, our methodology can also be used for inflow surfaces such as black holes. Although black holes are negligibly small in many astrophysical systems, their size can be significant in cases such as the tidal disruption of a star by a supermassive black hole (SMBH). If the SMBH is modeled as a point particle, it can accrete gas which exerts an (unphysical) outward pressure (Spaulding \& Chang, in preparation). Modeling the SMBH using an inflow boundary avoids this problem as the accreted gas is removed from the simulation entirely.

\section*{Acknowledgements}

We thank Philip Chang for useful and detailed discussions. We are supported by the NASA Astrophysical Theory Program (ATP) through NASA grant NNH17ZDA001N-ATP. We used the Extreme Science and Engineering Discovery Environment (XSEDE), which is supported by National Science Foundation (NSF) grant No. ACI-1053575. We acknowledge the Texas Advanced Computing Center (TACC) at The University of Texas at Austin for providing HPC resources that have contributed to the research results reported within this paper (URL: \url{http://www.tacc.utexas.edu}). We use the {\sevensize YT} software platform for the analysis of the data and generation of plots in this work \citep{yt}.

Computational resources supporting this work were also provided by the NASA High-End Computing (HEC) Program through the NASA Advanced Supercomputing (NAS) Division at the Ames Research Center. This material is based upon work supported by NASA under Award No. RFP19\_7.0 issued through the Wisconsin Space Grant Consortium and the National Space Grant College and Fellowship Program. Any opinions, findings and conclusions or recommendations expressed in this material are those of the author and do not necessarily reflect the views of the National Aeronautics and Space Administration.




\bibliographystyle{mnras}
\bibliography{references}


\bsp	
\label{lastpage}
\end{document}